# Multi-channel Absorption of Photons at Energies above 1 TeV

A. Subramanian[*]

5, Norton second street, Chennai- 600028

INDIA

It is shown that the absorption of photons at energies > 1 TeV (much higher than the mass of the Higgs boson ~ 100 GeV) is a multi-channel one as opposed to the purely electron pair like absorption at lower energies. The observation on muons and gamma rays from Cygnus X-3 point source at these energies (1 TeV and 10 TeV) is quantitatively accounted for. The expected multi-channel cross-sections of photons in air as a function of energy is given both for Coulomb dissociation and nuclear absorption upto limiting energies of $10^9$ GeV.

[*]Previously at the Tata Institute of Fundamental Research, Mumbai(Bombay)-400005 INDIA.
E-mail : ak_subra@yahoo.com

It is an ancient theoretical view that the absorption and emission of photons in matter proceeds through their couplings to charge oscillators in vacuum [1]. The physical materialization of charge particle pairs in the absorption process at high energies occurs via either nuclear Coulomb dissociation (the so called Bethe-Heitler process ) or nuclear photo-absorption.The coupling of the oscillators to photons is proportional to their charge to mass ratio (e/m ). charged elementary particles come with a variety of masses. The electron is the lightest among them, way down in mass compared to any other charged particle including the u and d quarks, so that the photon coupling is a single channel affair at photon energies we are familiar with (< 1 TeV ). There is purely electron –positron pair production in Coulomb dissociation in air ~ 500 mb. The quark pair (u,d) couplings are relatively small resulting in a small photo-nuclear absorption ~ 1 mb  in air. The current view is that most of the mass of an elementary  particle  comes from its coupling to the proposed Higgs boson field, the mass of which is estimated to lie around 100 GeV [2]. The coupling is Yukawa like for the fermions with a coupling constant proportional to the mass of the fermion itself. The range of this force is given by (m Higgs )$^{-1}$, an idea we are familiar with since the original proposal of Yukawa in the case of the pion field of the nucleon.

The interesting observation we now make is that for photon energies >> m Higgs  i.e., $\lambda$ photon << r Higgs, the Higgs boson field does not coherently oscillate with the charged fermion oscillators. There is a decoupling of the Higgs boson field for it is a spectator for these small length oscillators. The result is a coupling of "bare" masses of oscillators to the photon. The charge to mass ratio (e/m ) increases drastically and approaches infinity for bare mass tending to zero. All the elementary charged particles acquire the same order of bare mass which is zeroish and the photon statistically couples to all the available oscillators with a weight which is simply proportional to the square of their electrical charges ( $e^2$ ) in the infinite coupling limit. This situation was envisaged by us sometime ago [3]. The photon absorption at these energies and higher becomes multi-channel one, far removed from the situation where the photon is 100 % coupled to the electron pair at energies < 1 TeV.

If the above argument for the decoupling of the Higgs field at small distances is not convincing, one can take recourse to the possibility of the excitation of the Higgs vacuum [ 4, 3 ] in a limited domain of space with vacuum expectation value of the Higgs field = 0, turning all particle mass to zero. The scale at which this effect seems to happen is indicated from experiments to be  1 TeV, as shown below.     We have evidence for this complex  phase transition of the photon at  energies  ~ 1 TeV and > 10 TeV in two different experiments [5,6]. These are muon observations deep underground (or rock ) with minimum energies of 600 GeV and 5 TeV to penetrate the rock overburden in the two cases respectively. The muons pulsated with a periodicity of 4.8 hrs and came from the direction of Cygnus X-3,a stellar source. The astonishing observation in these two cases is that in the first one (muon energy > 600 GeV and parent photon energy > 1.2 TeV ), the ratio of flux of muons to parent gamma ray flux is 90%. In the second experiment with muon energy >5 TeV and parent gamma ray energy > 10TeV, the ratio is 34 +_ 5% . The gamma ray fluxes from Cygnus X-3 source which pulsates with a periodicity of 4.8 hrs is well measured in other atmospheric air shower experiments and a fitted integral energy spectrum in units of $cm^{-2} sec^{-1}$ is given [7] as

$$N(>E) = (2.5 +\_ 0.9)\ 10^{-7}\ E^{-1.128+/-0.013} \qquad (E\ in\ GeV)$$

for a broad energy range of 100 GeV – $10^4$ TeV. Parent gamma ray energies in the two experiments are gauged from the suspicion that muon pairs are produced in Coulomb dissociations in air, since in our view with the decoupling of the Higgs boson presumably at energies > 10 m Higgs = 1 TeV, the bare mass of the muon and electron (also the heavy tau-lepton, nay perhaps the W boson itself) is the same : zeroish.

The question of "bare" masses of elementary particles is a subtle one. We have QED, QCD and QFD (Higgs boson). If QFD is switched off with the decoupling of the Higgs boson or local excitation of its vacuum [4], the residual bare bare masses come from QED for the leptons and W but more importantly from QCD also for the quarks. At energies (< 1 TeV ), QCD masses of light quarks u and d must be large compared to the mass of the electron such that Coulomb dissociation of $pi^+pi^-$ is a closed channel. It was first pointed out by us [8] that at energies > 10 TeV, u and perhaps d quark masses become zeroish and plunge below the mass of the electron such that Coulomb dissociation of hadrons dominates the absorption cross-section. Geometric photo- nuclear cross-section was also envisaged in addition.

In the present situation, we are able to calculate the mu/gamma ratio of flux of 90% at ~ 1 TeV gamma ray energy only if we assume that pion channel (u,d quark channels ) is closed at this energy. A ratio of 46% results if pion channel was open. It is easy to see how a ratio of 90% could arise if only e, mu channels were open ( tau and W are too heavy for Coulomb dissociation here). We have to take into the calculation the fact that muon secondary energies from parent gamma energy E range merely E/2 +-20% due to non-relativistic c.m. mass of mu-pair in Coulomb dissociation. The conclusion we draw from this is that m(QCD) of u,d quarks >> m (electron) at 1 TeV.

The QCD situation is eased at the > 10 TeV observation of gamma rays in the second experiment [6]. Here we calculate a mu to gamma flux ratio (mu > 5 TeV and gamma > 10 TeV) of 33% if all the quark, lepton and W channels are open. The experimental figure for this ratio is 34+-5% which may be construed to be in good agreement. Thus we seem to substantiate the earlier speculation [8] that only at > 10 TeV QCD masses vanish , while we have reason to believe that QFD (Higgs boson) masses vanish at > 1 TeV. An inference is that Higgs boson does not couple appreciably to u,d quarks. Elsewhere we consider the question of strong coupling of Higgs boson to heavy particles and present details of the calculations there. We infer that ultimately the residual QED masses of charged particles are tiny (zeroish,<< mass of electron ).

As a summary of the present paper, we give in Figs 1a and 1b the the expected absorption cross-sections into various channels as a function of photon energy. The pictures are very strange and striking, far removed from our experiences at energies below 1 TeV.

The new situation is not easy for verification at Tevatron energies (limit 1 TeV) but at LHC energies (limit 7 TeV), we expect to see the striking effect of equal e-pair and mu-pair production. In a separate note we present some cosmic ray evidence to indicate that the second QCD phase transition actually occurs at an energy ~ 6 TeV, bringing in the multi-channel high photon cross-sections for hadro-production within the verification range of LHC energies.

The author wishes to thank his brother, A Gopalan for his immense help in computerization of this paper with the figures.

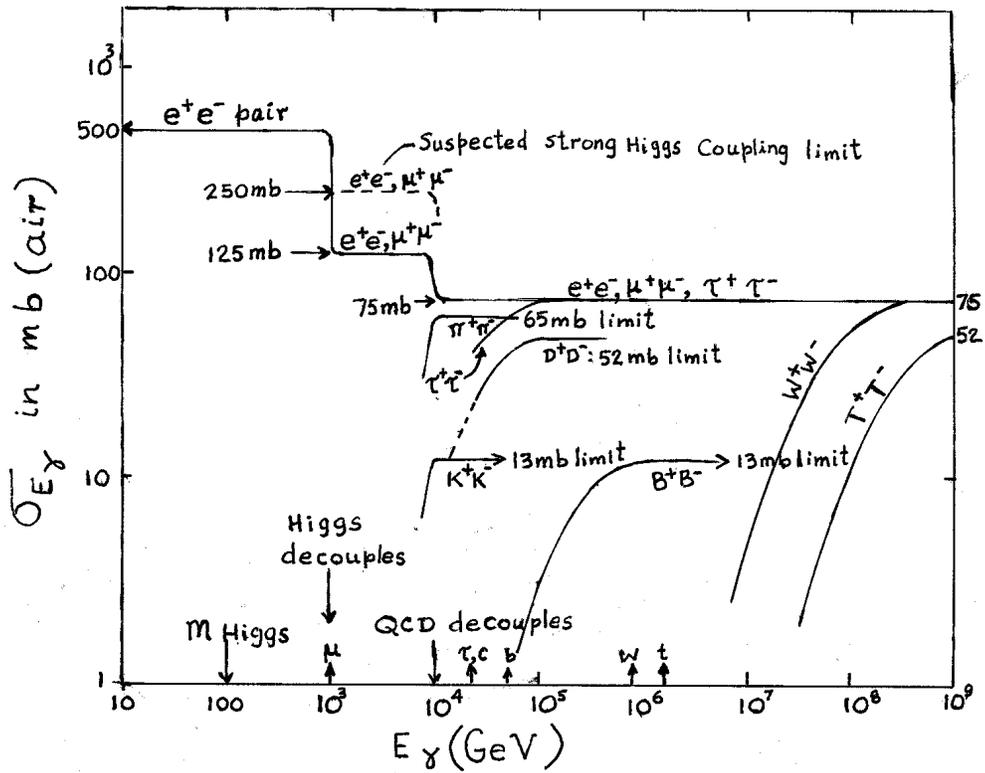

**Figure Captions**

**Fig. 1 a.**
Coulomb dissociation cross-sections in air as a function of gamma ray energy. The shifts in the transition curves are due to different longitudinal momentum transfers in the nuclear Coulomb field for different masses of particles. The limiting branching fractions are taken from ref. [3] . Small vertical arrows with particle nomenclatures on the energy scale are indicative of a possible scale in decoupling energies of the Higgs boson (coupling proportional to mass of particle).

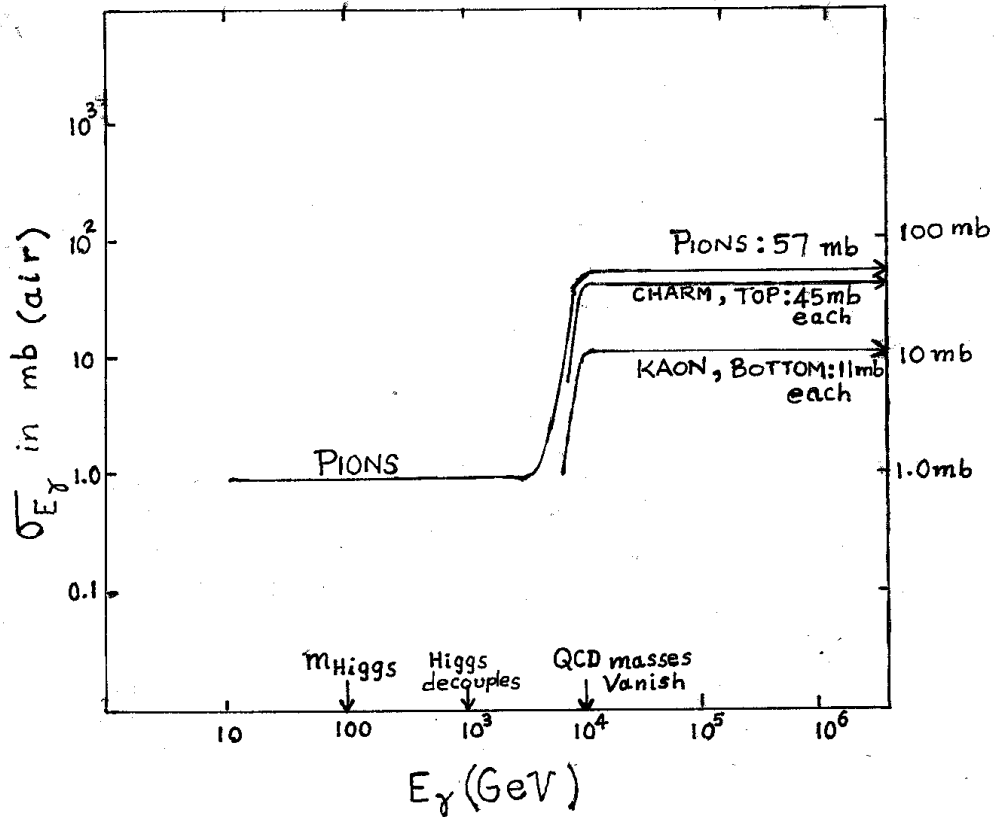

**Fig. 1 b.**

Photo-nuclear cross-sections in air for different hadrons reaching flat top at 10 TeV at which energy QCD masses of quarks are supposed to vanish. The flat top branching fractions are again taken from ref. [3]. The possible complication due to Higgs boson decoupling for heavy masses at different energies indicated in Fig. 1 a is not considered. This need not be considered if the effect is due to Lee and Wick's [4] vacuum excitation. The large flat top cross-sections suggested here are not in doubt at 10 TeV of incident energy of photon since longitudinal momentum  momentum transfers in gluonic field are small << 100 MeV /c ( peripheral production envisaged ) for all quarks except the top for which the flat top will set in only at an incident energy ~ 500 TeV ( lower than the Higgs decoupling energy of 1600 TeV, marked in Fig. 1a ). The QCD phase transition is likely to occur at ~ 6 TeV , as shown in a separate note. This will make the verification of the large multi-channel photo-nuclear cross-sections feasible at LHC.